\documentclass[aps,prb,twocolumn,floatfix,showpacs]{revtex4-1}
\usepackage{graphicx}
\usepackage{amssymb}
\usepackage{amsmath}
\usepackage{epstopdf}

\usepackage[pdftex]{color}

\newcommand{\bs}[1]{{\boldsymbol{#1}}}

\begin{document}

\title{Coexistence of Ferromagnetism and Superconductivity\\in Noncentrosymmetric Materials with Cubic Symmetry}

\author{Titus Neupert} 
\affiliation{
Condensed Matter Theory Group, 
Paul Scherrer Institute, 5232 Villigen PSI, Switzerland}
            
\author{Manfred Sigrist}%
\affiliation{%
Institute for Theoretical Physics, ETH Z\"urich, 8093 Z\"urich, Switzerland
}%

\date{\today}

\begin{abstract}
This is a model study for the emergence of superconductivity in ferromagnetically ordered phases of cubic materials whose crystal structure lacks inversion symmetry. A Ginzburg-Landau-type theory is used to find the ferromagnetic state and to determine the coupling of magnetic order to superconductivity. It is found that noncentrosymmetricity evokes a helical magnetic phase. If the wavelength of the magnetic order is long enough, it gives rise to modulations of the order parameter of superconductivity, both in modulus and complex phase. 
At magnetic domain walls the nucleation of superconductivity is found to be suppressed as compared to the interior of ferromagnetic domains.
\end{abstract}

\maketitle

\section{Introduction}

If the unit cell of a three-dimensional crystal is noncentrosymmetric, i.e., not invariant under the parity operation, the spatial inversion symmetry is broken. Noncentrosymmetricity allows for the Dzyaloshinskii-Moriya interaction that entails the breaking of spin rotation symmetry due to antisymmetric spin-orbit coupling (ASOC).
Many noncentrosymmetric material feature unusual properties, when ordering phenomena such as superconductivity or magnetic order, that break other symmetries in addition to inversion and spin rotation, are present.
For example, ASOC and magnetic order lead to a helical magnetic structure in MnSi~\cite{BAK80} and Ba$_2$CuGe$_2$O$_7$~\cite{ZHE96}, where they give rise to magnetic field-induced ferroelectricity~\cite{MUR10}.

Superconductivity with a noncentrosymmetric crystal structure generically appears in a mixed parity state \cite{SIG07}. 
In addition, many of the known noncentrosymmetric superconductors, such as CePt$_3$Si~\cite{BAU04,SA04}, CeRhSi$_3$~\cite{KIM07-2}, and UIr~\cite{AKA06}, show states of magnetic order and in some regions of the phase diagram superconductivity coexists with the magnetic order. A simple consideration of the combined action of ASOC and  time-reversal symmetry breaking magnetic fields or magnetization on the energy bands reveals the possibility of spatial modulations of the order parameter of superconductivity (finite-$q$-pairing)~\cite{DIM03,KAUR05,AG07,MI09}.

The aim of this paper is to study noncentrosymmetric systems which show coexistence of magnetic order and superconductivity. To be concrete, we will focus on a cubic crystal without inversion center described by the point group $ O $ and restrict our study to the case of ferromagnetic order, where the wavelength of modulations of the magnetization is much longer than the lattice spacing. We will use the generalized Ginzburg-Landau-approach.  In sec.~II we determine the nature of the helical ferromagnetic phase in presence of ASOC. In sec.~III we subsequently study, how the superconducting state is altered on the background of this magnetic order. We find that noncentrosymmetricity causes the magnetic moment to follow a helical modulation and introduces a new length scale for the superconducting order parameter.
Depending on the ratio of the magnetic wavelength and this length scale, the order parameter of superconductivity either remains homogeneous or exhibits a modulation both in complex phase and absolute value in this magnetic phase. 
Finally, in sec.~IV we consider a limit where the magnetic state resembles a filamentary domain structure. We show that superconductivity will nucleate in the interior of the domains rather than at the domain walls.

\section{Magnetic State}
In order to determine the magnetic state, the free energy density is expanded in the magnetization $\boldsymbol{m}(\boldsymbol{r})$ as a three-dimensional order parameter. The expression must be invariant under time reversal symmetry and under the symmetry transformations of the cubic point group $O$. Spin-orbit coupling ties rotations of spin and orbital degrees of freedom together, such that the free energy has to be invariant under a simultaneous rotation in both spin and orbital space. The magnetization $\boldsymbol{m}(\boldsymbol{r})$ belongs to the irreducible (vector) representation $\Gamma_4$ of the point group $O$.
The second-order terms of the free energy read
\begin{eqnarray}
\begin{split}
F_{\text{M}}^{(2)}:=&\,
\int{d^3r} \Bigl\{\alpha\boldsymbol{m}^2+\frac{\tau_0-\tau_1}{2}\left(\nabla\times\boldsymbol{m}\right)^2+\tau_2\left(\nabla\cdot\boldsymbol{m}\right)^2
\\
&+
\tau_3\sum_{i,j}\left[3\delta_{ij}(\partial_im_i)^2-(\partial_im_i)(\partial_jm_j)\right]\\
&+
\frac{\tau_0+\tau_1}{4}\sum_{i\neq j}\left(\partial_im_j+\partial_jm_i\right)^2+\vartheta\boldsymbol{m}\cdot\left(\nabla\times \boldsymbol{m}\right)\Bigr\},
\end{split}
\label{FunMagCubic}
\end{eqnarray}
where $\vartheta$ and $\tau_i\ (i=0\ldots3)$ are temperature-independent phenomenological parameters and $\alpha$ has the usual temperature dependence $\alpha=\alpha'(T-T^{\ }_{\text{M}})$, with $\alpha'>0$. Here, $T^{\ }_{\text{M}}$ is the transition temperature for a state of homogeneous magnetization. The effect of noncentrosymmetricity manifests itself in the presence of a Lifschitz-invariant term proportional to $\vartheta$~\cite{MI09,SAM04}, which would not be allowed had we considered the centrosymmetric cubic point group $O_h$. We demand that $F_{\text{M}}^{(2)}$ is bound from below towards arbitrarily strong modulations of $\boldsymbol{m}(\boldsymbol{r})$. Amongst others this leads to the condition
$\tau_0>0$.

\begin{subequations}
The vector field $\boldsymbol{m}(\boldsymbol{r})$ that minimizes $F_{\text{M}}^{(2)}$ is given by a helix with a wavevector $\boldsymbol{k}_0$. Depending on the parameters $\tau_i\ (i=0\ldots3)$, $\boldsymbol{k}_0$ is either aligned parallel to a coordinate axis (6-fold degenerate) or parallel to a body diagonal (8-fold degenerate).
Let us assume $\boldsymbol{k}_0=k_0\boldsymbol{e}_z$ from here on.
In that case, the magnetization takes the form
\begin{equation}
\boldsymbol{m}(\boldsymbol{r})=m_0\left[\sin(\pm k_0z)\boldsymbol{e}_x+\cos\left(k_0z\right)\boldsymbol{e}_y\right].
\label{MagCubic}
\end{equation}
The two signs correspond to two energetically degenerate chiralities and $k_0=|\vartheta|/(2\tau_0)$. The magnitude $m_0$ of the magnetization would be determined by  terms of the order $\boldsymbol{m}^4$. For the time being we ignore these terms and consider $ m_0 $ as a parameter. 
The transition temperature to the magnetic state is given by
\begin{equation}
T^*_{\text{M}}=T^{\ }_{\text{M}}+\frac{\vartheta^2}{4\alpha'\tau_0}
\end{equation}
and is always larger than $T^{\ }_{\text{M}}$, owing to $\tau_0>0$. In the limit of a centrosymmetric system ($\vartheta\rightarrow0$) we recover a state of homogeneous magnetization with the corresponding transition temperature, $T^{\ }_{\text{M}}$.
\end{subequations}

\section{Superconductivity in the magnetic phase}

In this section, we study the influence of homogeneous and helical magnetic order as given by eq.~\eqref{MagCubic} on the superconducting order. We assume the temperature to be sufficiently below the transition temperature to the magnetic phase, such that an emergent superconducting order parameter does not change the magnetization considerably. For simplicity, a complex scalar superconducting order parameter $\eta(\boldsymbol{r})$ belonging to the $\Gamma_1$ representation of the point group $O$ is considered. 
The expansion of the free energy up to the order $\eta^2$ reads
\begin{equation}
\begin{split}
F^{\ }_{\text{SC}}:=&\,\int{d^3r}\Bigl\{a\left|\eta\right|^2+b\left|\bs{D}\eta\right|^2+
u\boldsymbol{m}^2\left|\eta\right|^2+\\
& \hphantom{\int{d^3r}\Bigl\{} i v\left(\eta^*\boldsymbol{m}\cdot\bs{D}\eta-\eta\,\boldsymbol{m}\cdot\bs{D}^*\eta^*\right)
\Bigr\},
\end{split}
\label{FunScCubic}
\end{equation}
where $\bs{D}=\nabla-2ie\bs{A}/(\hbar c)$ with $\bs{A}$ being the vector potential.
Here, $b$, $u$ and $v$ are temperature-independent phenomenological parameters and $a$ has linear temperature dependence $a=a'(T-T^{\ }_{\text{SC}})$ with $a'>0$. $T^{\ }_{\text{SC}}$ denotes the critical temperature of superconductivity for $\boldsymbol{m}=0$. 
Demanding that $F^{\ }_{\text{SC}}$ is bound form below if $\eta$ has strong spatial fluctuations necessitates $b>0$. The term proportional to $u$ represents the paramagnetic depairing effects of the magnetization on the Cooper-pair formation, thus we assume $u>0$. Noncentrosymmetricity is again reflected by the presence of a Lifschitz-term proportional to $v$, which is forbidden in case with inversion symmetry. 
Note that the Lifschitz-term introduces a new length-scale $|\xi_{\text{L}}|$ for the superconducting order, where $\xi_{\text{L}}:=b/(vm_0)$ and is not singular at the transition to the superconducting phase.
In the following we shall assume the limit of strong type-II superconductivity
in the sense that the length scales $k^{-1}_0$, $\xi_{\text{L}}$, and the coherence length $\sqrt{b/a}$ are assumed to be much smaller than the magnetic penetration depth $\lambda$.~\cite{MI09} In this limit, we can neglect the effect of the vector potential $\bs{A}$ and replace $\bs{D}\to \nabla$ in eq.~\eqref{FunScCubic}.

\subsection{Homogeneous Magnetization}

\begin{subequations}
Before addressing the helical magnetic order, we first consider the effect of a homogeneous magnetization $\boldsymbol{m}=m_0\boldsymbol{e}_y$ on the onset of superconductivity, i.e., we take the limit $k_0^{-1}\to \infty$. Minimizing the functional~\eqref{FunScCubic} with respect to $\eta$ straightforwardly yields a modulation of the order parameter as
\begin{equation}
\eta(x)=\eta_0 \text{e}^{i y/\xi_{\text{L}}}.
\end{equation}
The transition temperature is then given by
\begin{equation}
T^{(1)}_{\text{SC}}=T^{\text{hom}}_{\text{SC}}+\frac{b}{a'\xi^2_{\text{L}}}
\label{T1}
\end{equation}
and shows the advantage of the phase modulation of the order parameter as compared to the transition temperature to a homogeneous superconducting state $T^{\text{hom}}_{\text{SC}}=T^{\ }_{\text{SC}}-um_0^2/a'$.
The appearance of these phase modulations of $\eta$ in noncentrosymmetric superconductors in a homogeneous magnetic field was already pointed out in various other studies \cite{ED89,DIM03,KAUR05,AG07,MI09}.
\end{subequations}

\subsection{Helical Magnetization}

We now turn to the more subtle effect of the helical magnetization given by eq.~\eqref{MagCubic} on the superconducting state. The variation of $F^{\ }_{\text{SC}}$ with respect to $\eta$ yields
\begin{subequations}
\begin{equation}
0=\left[\partial_\zeta^2+A(B)+2B\cos(2\zeta)\right]\eta_{k,\varphi}(\zeta),
\label{Mathieu}
\end{equation}
where a Fourier transformation of the superconducting order parameter in the coordinates $x$ and $y$ was performed and $(k_x,k_y)=(k\cos\varphi,k\sin\varphi)$.

We introduce here the parameters
\begin{eqnarray}
A(B)&:=&\frac{4a'(T^{\text{hom}}_{\text{SC}}-T)}{bk_0^2}-\left(\frac{B\xi_{\text{L}}k_0}{2}\right)^2 \label{A},\\
B&:=&\frac{4k}{\xi_{\text{L}}k_0^2}\label{B},
\end{eqnarray}
and the $z$ coordinate is substituted by $\zeta$ as
$\zeta:=\left(k_0z-\varphi-\pi/2\right)/2$.
\end{subequations}
	
\begin{figure}[t]
			\includegraphics[width=0.450\textwidth]{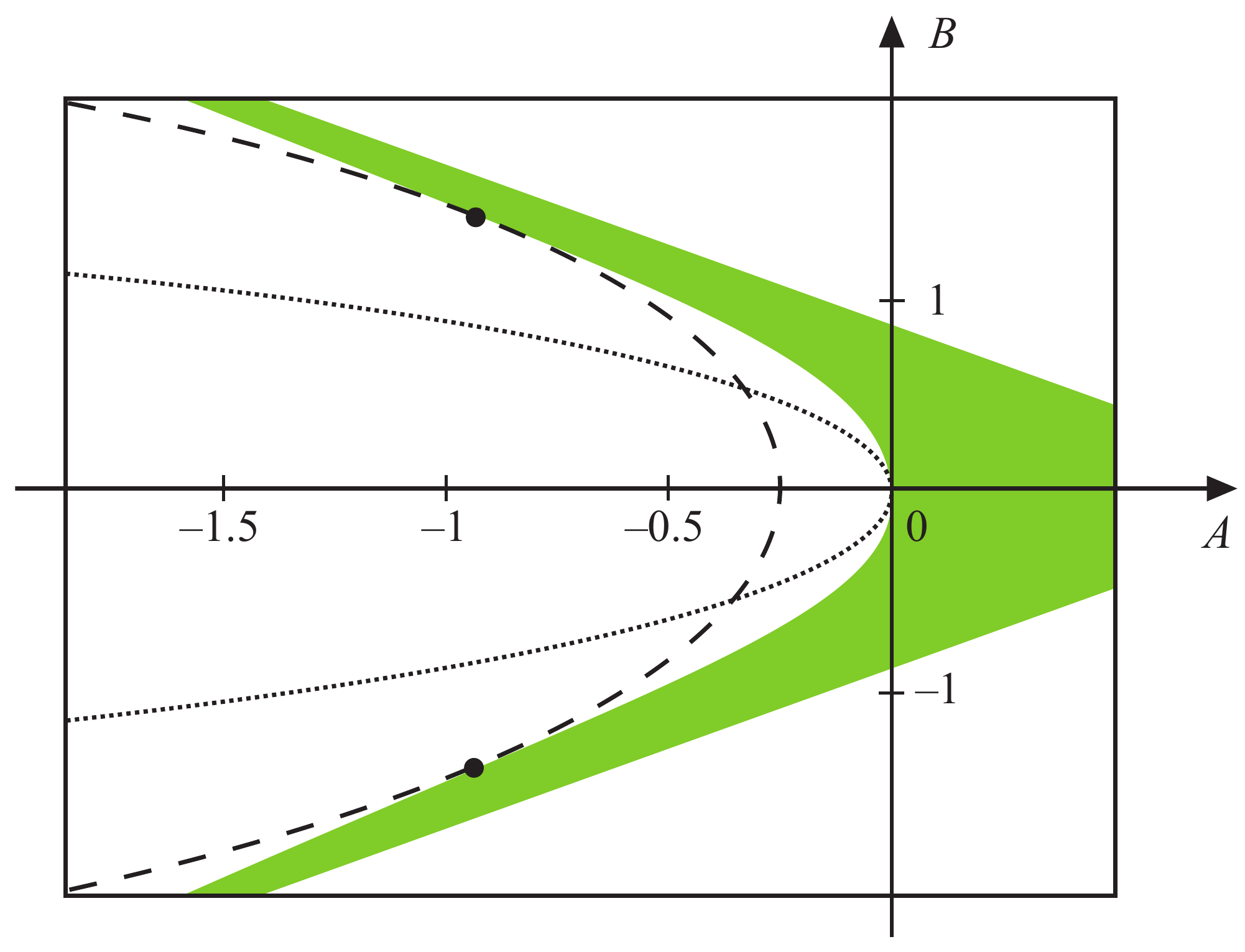}
			\caption{\label{fig:Mat} (Color online) Portion of the stability diagram of Mathieu's equation. Equation~\eqref{A} defines a parabola $A_T(B)$. As the temperature decreases, the parabola migrates from left to right. The transition temperature is reached when the parabola touches the region of stability (dots on dashed curve) at some critical value $B_{\text{c}}$, which in turn determines $k$. If the curvature of the parabola at $k=0$ is larger than that of the stability region (dotted curve), there will be no contact with the region of stability until the transition to the homogeneous state $B_{\text{c}}=0$.}
\end{figure}

In eq.~\eqref{Mathieu} we identify Mathieu's differential equation with the variable $\zeta$. Mathieu's equation cannot be solved analytically in a closed form. However, one can identify a region in the the $A$-$B$-parameter space where the solutions of eq.~\eqref{Mathieu} are bounded for all $\zeta\in\mathbb{R}$. This stability region is displayed in fig.~\ref{fig:Mat} as a shaded area. For parameter values $(A,B)$ inside this stability region, the system would be in a stable superconducting state. 
Equation~\eqref{A} defines a parabola $A_T(B)$ of energetically degenerate parameter values in the $A$-$B$-plane. These parabolas can be labeled by temperature $T$ via the temperature dependence of $A$. For sufficiently high temperatures these parabolas do not intersect with the region of stability. As the temperature is lowered, the superconducting instability occurs when the parabola $A_T(B)$ touches first the boundary of the  (shaded) stability region in fig.\ref{fig:Mat}.
Different types of solutions depend on the curvature of the parabola around $ B =0 $ which should be compared with that of the boundary to the stability region, which can expanded at $ B=0 $ to $A(B)\approx-B^2/2+7B^4/128$~\cite{AB64}. If the curvature of $A_T(B)$ is larger than that of the stability region, corresponding to the condition
\begin{equation}
|\xi_{\text{L}}|k_0 > 
\sqrt{2} ,
\label{CurvCond}
\end{equation}
then the touching point is at $B_{\text{c}}=0$ (see dotted line in Fig.~\ref{fig:Mat}) such that with $ k=0 $ the superconducting order parameter is homogeneous. For the condition opposite to eq.~\eqref{CurvCond} the touching point is at finite values of $ B_c \approx \pm \{16 (2 - \xi_L^2 k_0^2)/7 \}^{1/2}  $ ($ \ll 1$) yielding a modulated order parameter with a finite $k$, as shown by the dashed line in fig.~\ref{fig:Mat}. 
Equation~\eqref{CurvCond} can be seen as an analogue to the condition on the Ginzburg-Landau parameter $\kappa$ that appears in the discussion of the vortex phase of superconductors. As in our case, we compare a superconducting length scale ($\xi_{\text{L}}$) with a magnetic length scale ($k_0^{-1}$) and obtain an inhomogeneous superconducting state if the superconducting length scale is shorter.

\begin{figure}[t]
			\includegraphics[width=0.45\textwidth]{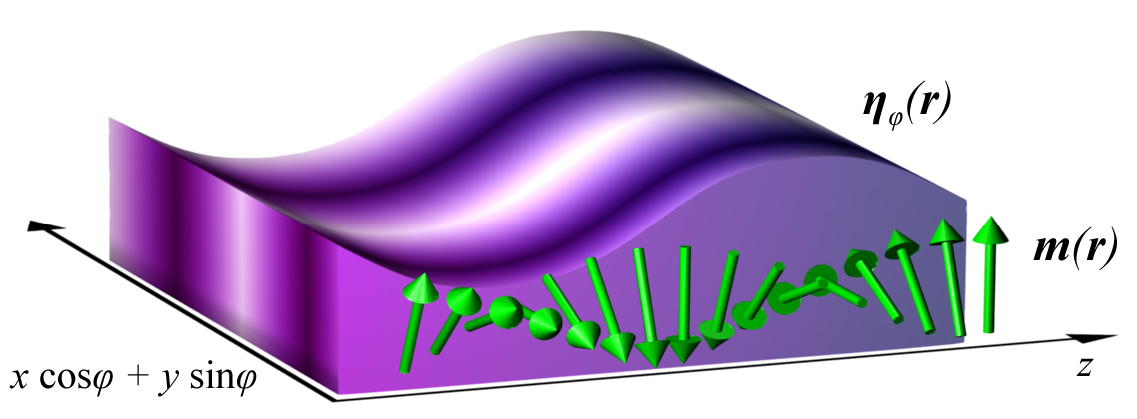}
			\caption{\label{fig:order parameter} (Color online) Schematic picture of the superconductivity order parameter $\eta_\varphi(\boldsymbol{r})$ given by eq.~\eqref{sol} that nucleates in the presence
of the sketched helical magnetization. It exhibits an amplitude modulation in the $z$-direction, and in addition to that
features a phase modulation perpendicular to the $z$-direction (depicted by the colour gradient).}
\end{figure}

For the inhomogeneous case, the solution features a continuous circular degeneracy in the $k_x$-$k_y$-plane, parameterized by $\varphi$. The elementary solution is given in terms of the lowest order even Mathieu function~$\text{ce}_0$ \cite{AB64} by
\begin{equation}
\begin{split}
\eta_\varphi(\boldsymbol{r})=&\,\eta^{\ }_0 \,\text{exp}\left[ik_0^2\xi_{\text{L}}B_{\text{c}}(x\cos\varphi+y\sin\varphi)\right]\\
&\times\text{ce}_0\left(k_0z-\varphi-\pi/2,B_{\text{c}}\right).
\end{split}
\label{sol}
\end{equation}
This form of the order parameter is sketched in fig.~\ref{fig:order parameter} together with the helical magnetic state.
The general solution is a superposition $\eta^{\ }_n:=\sum_{l=1}^n c^{\ }_l\eta^{\ }_{\varphi_l}$ with the complex-valued coefficients $c^{\ }_l$. To find the set of coefficients $c_l$ and phases $\varphi_l$ that minimizes the free energy, we use Abrikosov's parameter $\beta$ defined by 
\begin{equation}
\beta:=\frac{\left\langle|\eta|^4\right\rangle}{\left\langle|\eta|^2\right\rangle^2},
\label{eq:Abrikosov parameter}
\end{equation}
where $\left\langle \ldots\right\rangle$ is the spatial average. The solution that minimizes $\beta$ is realized. A minimum requires that the partial derivatives $\partial_{\varphi_l}\beta$ and $\partial_{c_l}\beta$ vanish for all $l$. For the phases, this yields the condition $\varphi_i-\varphi_j=\pi N_{ij}/2,\ \forall i,j$ with some $N_{ij}\in \mathbb{Z}$. An explicit evaluation of $\beta$ for the remaining cases $n=1\ldots4$ reveals the optimal solution to be
\begin{equation}
\eta_{\varphi}^{\text{opt},\pm}=\eta_{\varphi}+i\eta_{\varphi\pm\pi/2}.
\label{eq:optimal superposition}
\end{equation}
From this result we see that the order parameter acquires a long-wavelength phase and amplitude modulation perpendicular to the wavevector of the helical magnetic order $\boldsymbol{k}_0$ and an amplitude modulation with the same wavevector as the magnetic order. To lowest order in $ B_c^2 \propto (2-k_0^2\xi_{\text{L}}^2)$, the transition temperature of this inhomogeneous superconducting state is given by
\begin{subequations}
\begin{equation}
T^{(2)}_{\text{SC}}=T^{\text{hom}}_{\text{SC}}+\frac{bk_0^2}{14a'}\left(2-k_0^2\xi_{\text{L}}^2\right)^2,
\end{equation}
and the superconducting order parameter has the approximate form
\begin{equation} \begin{array}{ll}
\eta^{\text{opt},\pm}_{\varphi =0} (\boldsymbol{r})  \approx &  \frac{\eta^{\ }_0}{\sqrt{2} } 
\left\{ e^{i \sqrt{2} B_c k_0 x } \left[1 + \frac{B_c}{2} \cos(2k_0 z ) \right] \right. \\ 
& \left. + i e^{\pm i \sqrt{2} B_c k_0 y } \left[1 - \frac{B_c}{2} \cos(2k_0 z) \right] \right\},
\end{array}
\end{equation}
taking $ 0 < B_c \ll 1 $ and $ \xi_L k_0 \approx \sqrt{2} $. 
\end{subequations}									

\section{Magnetic solitons and Superconductivity}

The analysis of the free energy $F_{\text{M}}^{(2)}$ showed that ASOC favors an inhomogeneous magnetic state. So far we ignored in the discussion the explicit form of the fourth order terms in the free energy expansion. Besides fixing the magnitude of the magnetic moment, these terms may also introduce additional features such as crystal anisotropy of the magnetic moments. The fourth order terms allowed within the point group $ O $, ignoring any gradient terms, read
\begin{equation}
F_{\text{M}}^{(4)}:=\int{d^3r} 
\left(
\frac{3}{2}\beta_0 \boldsymbol{m}^4+
2\beta_1\sum_{i}m_i^4
\right).
\label{FunMagCubic4}
\end{equation}
Here, $\beta_0$ and $\beta_1$ are temperature-independent phenomenological parameters. The expression shows that not all directions of the magnetization are degenerate in energy. If we simply insert the helical solution~\eqref{MagCubic} we obtain
\begin{equation}
F_{\text{M}}^{(4)}= \,m_0^4\int{d^3r} \left[\text{const.}+\frac{\beta_1}{2}\cos(4k_0z)\right],
\end{equation}
which does not minimize this part of the free energy. 
It is instructive to solve for the magnetization $\bs{m}(\bs{r})$, that minimizes the total magnetic free energy $F_{\text{M}}^{(2)}+F_{\text{M}}^{(4)}$, in order to understand the effect of the anisotropy term $\beta_1$ qualitatively. As before, we assume that the magnetization has no component in $\bs{e}_z$-direction and will be a function of $z$ only. With the complex notation $M(z)=m_x(z)+i m_y(z)$, the variational equation reads
\begin{subequations}
\begin{equation}
\begin{split}
0=&
\left(
-\tau_0^{\ }\partial_z^{2}+i\vartheta\partial_z^{\ } +\alpha
\right)
M(z)\\
&
+
3(\beta_0+\beta_1)
\left|M(z)\right|^2 M(z)
+\beta_1\bar{M}^3(z).
\end{split}
\label{eq:}
\end{equation}
To linear order in $\beta_1$, the solution is given by
\begin{equation}
M(z)=m_0e^{i k_0 z}\left[1+\frac{\beta_1}{2\beta_0} 
\frac{\vartheta^2-4 \alpha\tau_0}{9\vartheta^2-4 \alpha\tau_0}e^{-4 i k_0 z}\right]
\label{eq:}
\end{equation}
where now $m_0^2=\alpha'(T^*_{\text{M}}-T)/[3(\beta_0+\beta_1)]$. 
Thus, the anisotropy term pins the magnetization 
$\bs{m}(\bs{r})$ parallel to $\bs{e}_x$ or $\bs{e}_y$ (parallel to $\bs{e}_x\pm\bs{e}_y$) for $\beta_1<0$ (for $\beta_1>0$), besides a modulation of the amplitude $|\bs{m}(\bs{r})|=|M(z)|$.
In the limit of a strong anisotropy, the magnetic state could be viewed as parallel planes of magnetic solitons, where each soliton twists the magnetization-vector by $90^{\circ}$.
\end{subequations}

We will now address the question, how superconductivity nucleates in the presence of such a filamentary magnetic structure.
From our analysis in the previous section, we know that in the limit $|\xi_{\text{L}}|k_0 > \sqrt{2}$ a homogeneous superconducting order nucleates despite the modulated magnetic background.
In the opposite case, $|\xi_{\text{L}}|k_0 < \sqrt{2} $, for which a spatially modulated superconducting order is found, first of all the solitons in the magnetization lift the continuous degeneracy of the solution~\eqref{eq:optimal superposition} parameterized by $\varphi$. 

\begin{subequations}
Then, the question arises whether the superconducting order parameter nucleates at the soliton (domain wall) position or rather in the interior of the domains.
To address this, let us consider the nucleation of bound states at an isolated domain wall at which the magnetization is tilted from $\bs{m}(\bs{r})\parallel \bs{e}_x$ for $z\to-\infty$ to $\bs{m}(\bs{r})\parallel \bs{e}_y$ for $z\to\infty$. We shall for the moment assume that the amplitude $|\bs{m}(\bs{r})|\equiv|M(z)|=m_0$ is constant across the domain wall. The magnetization can thus be parametrized by a single function $\theta(z)=-\theta(-z)$ with $\theta(z\to\pm\infty)=\pm\pi/4$ as
\begin{eqnarray}
M(z)&=&m_0\,\mathrm{exp}\left[i\theta(z)+i\pi/4\right].
\label{MagCubicSoliton}
\end{eqnarray}
Insertion of the magnetization~\eqref{MagCubicSoliton} in the free energy for the superconducting order parameter, Eq.~\eqref{FunScCubic}, yields upon Fourier transformation in the $x$ and $y$ coordinates the variational equation
\begin{equation}
\left[-\partial_z^2+\frac{2k}{\xi_{\text{L}}}V(z)
\right]\eta^{\ }_{k}(z)
=
-\left[
\frac{a}{b}+\frac{um_0^2}{b}
+2k^2
\right]\eta^{\ }_{k}(z)
,
\label{eq:Schroedinger}
\end{equation}
where
\begin{equation}
V(z):=-\sin[\theta(z)+\pi/4]-\cos[\theta(z)+\pi/4].
\end{equation}
\end{subequations}
Here, we have chosen $k_x=k_y\equiv k$ in accordance with the symmetry of the problem and to obtain a binding potential.
Eq.~\eqref{eq:Schroedinger} is reminiscent of the one-dimensional Schr\"odinger equation with the potential $V(z)$ and an energy eigenvalue given by the square bracket on the rhs. This analogy immediately delivers the inequality
\begin{subequations}
\begin{equation}
-\frac{2k}{\xi_{\text{L}}}\sqrt{2}>\frac{a}{b}+\frac{um_0^2}{b}
+2k^2,
\label{eq:}
\end{equation}
since the energy of the lowest bound state is always larger than the potential minimum $V(0)=-\sqrt{2}$. 
Via the temperature dependence of $a$, the energy eigenvalue of the lowest bound state determines the temperature at which the superconductivity nucleates at the domain wall. 
The transition temperature $T^{\mathrm{dw}}_{\mathrm{SC}}$ of the bound state therefore satisfies
\begin{equation}
T^{\mathrm{dw}}_{\mathrm{SC}}<T_{\mathrm{SC}}-\frac{um_0^2}{a'}+\frac{b}{a'\xi_{\text{L}}}=T^{(1)}_{\mathrm{SC}},
\label{eq:inequality trans temp}
\end{equation}
where $T^{(1)}_{\mathrm{SC}}$ is the bulk transition temperature defined in eq.~\eqref{T1}. This inequality holds independent of the concrete form of the function $\theta(z)$.
We conclude that superconductivity will nucleate in the interior of the domains rather than at the magnetic solitons (domain walls). 
\end{subequations}

This result relies on the assumption that $|\bs{m}(\bs{r})|$ is spatially constant. If we relax this assumption and consider the case in which the magnetization is suppressed near the domain wall to a value $m'_0<m_0$, the superconducting order near the domain wall would be less affected by the paramagnetic depairing effect, represented by the parameter $u$. The upper bound for $T^{\mathrm{dw}}_{\mathrm{SC}}$ in eq.~\eqref{eq:inequality trans temp} then exceeds $T^{(1)}_{\mathrm{SC}}$, thereby opening the way for an reversion of the inequality $T^{\mathrm{dw}}_{\mathrm{SC}}<T^{(1)}_{\mathrm{SC}}$.

\section{Conclusions}
In our study on the coexistence of ferromagnetism and superconductivity in noncentrosymmetric materials we found that ASOC, represented by Lifschitz-terms in the free energy expansions, gives rise to unusual modulations of the order parameters. Where a centrosymmetric material would have a homogeneously magnetized ferromagnetic ground state, an arbitrarily small parity violation generates a state of helical magnetization.

The possible modulations in the superconducting order parameter in presence of magnetic order were found to be governed by the ratio of two length scales, $|\xi_{\text{L}}|$ and the characteristic length of magnetic modulations $k_0^{-1}$. When the magnetic length scale is larger, the superconducting order is not homogeneous. This extends to the limit of homogeneous magnetization $k_0^{-1}\to\infty$, where a complex phase modulation was found.
In presence of the helical magnetic phase, a state of simultaneous complex phase and amplitude modulation develops, if the magnetic wavelength is large enough.

We closed our discussion with the consideration of a magnetic state of filamentary solitons, which is obtained for a strong anisotropy parameter in the magnetic free energy. We found that this will force the order parameter of superconductivity to nucleate with the same filamentary structure, creating a stripe-like state with maxima in between two solitons. 

The complex phase winding of the order parameter of non-centrosymmetric superconductors that are exposed to a magnetic field was noticed theoretically a long time ago~\cite{ED89}, but an experimental verification of this state is still lacking. The inhomogeneous superconducting states that we find in this work show both complex phase and amplitude modulations. The latter would entail an anisotropic resistivity drop at the criticality for current directions perpendicular and parallel to the wavevector of the modulation and are therefore more accessible to an experimental observation.
In fact, such anisotropic superconducting transitions were reported for the antiferromagnetically ordered noncentrosymmetric CeRhSi$_3$ in ref.~\onlinecite{KIM07-2}, where superconductivity and magnetism show and interesting interplay.

We would like to thank D.F. Agterberg, N. Hayashi, V. Mineev, Y. Yanase and S. Guerrero for helpful discussions. This work was financially supported by the Swiss National Science Foundation and the NCCR MaNEP.

\end{document}